\documentstyle[preprint,prl,aps,epsfig]{revtex}
\tightenlines

\newcommand{\kpitwog} {\mbox{$K^+ \! \rightarrow \! \pi^+ \pi^0 \gamma$}}

\newcommand{\kpitwo}    {\mbox{$K_{\pi 2}$}}
\newcommand{\kpitwogam}    {\mbox{$K_{\pi 2 \gamma}$}}

\newcommand{\chisq}    {\mbox{$\chi^2$}}

\begin{document}

\preprint{\vbox{\hbox{BNL-67407}
\hbox{PRINCETON/HEP/2000-3}
\hbox{TRI--PP--00--21}
\hbox{KEK Preprint 2000-16}}}

\title{Measurement of Direct Photon Emission in \kpitwog\ Decay}


\author{
S.~Adler$^1$, M.~Aoki$^5$\cite{masa}, M.~Ardebili$^4$, M.S.~Atiya$^1$,
P.C.~Bergbusch$^{5,7}$, E.W.~Blackmore$^5$, D.A.~Bryman$^{5,7}$,
I-H.~Chiang$^1$, M.R.~Convery$^4$, M.V.~Diwan$^1$, J.S.~Frank$^1$,
J.S.~Haggerty$^1$, T.~Inagaki$^2$, M.M.~Ito$^4$, S.~Kabe$^2$,
S.H.~Kettell$^1$, Y.~Kishi$^3$, P.~Kitching$^6$, M.~Kobayashi$^2$,
T.K.~Komatsubara$^2$, A.~Konaka$^5$, Y.~ Kuno$^2$, M.~Kuriki$^2$,
T.F.~Kycia$^1$\cite{TFK}, K.K.~Li$^1$, L.S.~Littenberg$^1$,
J.A.~Macdonald$^5$, R.A.~McPherson$^4$,
P.D.~Meyers$^4$, J.~Mildenberger$^5$, N.~Muramatsu$^2$, T.~Nakano$^3$,
T.~Numao$^5$, J.-M.~Poutissou$^5$, R.~Poutissou$^5$,
G.~Redlinger$^5$\cite{george},
T.~Sato$^2$, T.~Shinkawa$^2$, F.C.~Shoemaker$^4$,
R.~Soluk$^6$, J.R.~Stone$^4$, R.C.~Strand$^1$,
S.~Sugimoto$^2$, C.~Witzig$^1$, and Y.~Yoshimura$^2$
\\ (E787 Collaboration) }

\address{ 
$^1$Brookhaven National Laboratory, Upton, New York 11973\\
$^2$High Energy Accelerator Research Organization (KEK), 
Oho, Tsukuba, Ibaraki 305-0801, Japan \\ 
$^3$RCNP, Osaka University, 10--1 Mihogasaki, Ibaraki, Osaka 567-0047, 
Japan \\ 
$^4$Joseph Henry Laboratories, Princeton University, Princeton, 
New Jersey 08544 \\
$^5$ TRIUMF, 4004 Wesbrook Mall, Vancouver, British Columbia,
Canada, V6T 2A3\\
$^6$ Centre for Subatomic Research, University of Alberta, Edmonton,
Alberta, Canada, T6G 2N5\\
$^7$ Department of Physics and Astronomy, University of British Columbia, 
Vancouver, BC, Canada, V6T 1Z1
}


\maketitle

\begin{abstract}

We have performed a measurement of the $\kpitwog$ decay and have
observed $2 \times 10^4$ events.  The best fit to the decay spectrum
gives a branching ratio for direct photon emission of
$(4.7\pm0.8\pm0.3)\times 10^{-6}$ in the $\pi^+$ kinetic
energy region of 55 to 90 MeV and requires no component due to
interference with inner bremsstrahlung.
\end{abstract}


\draft

The radiative decay $\kpitwog$ ($\kpitwogam$) in which the $\gamma$ is
directly emitted (DE) is sensitive to important aspects of the low
energy hadronic interactions of mesons~\cite{DEtheory}.  
Although the DE component is difficult to observe
because it is very small compared to the dominant
inner bremsstrahlung (IB) associated
with the $K^+\! \rightarrow \! \pi^+ \pi^0$ ($\kpitwo$) decay,
it may be isolated kinematically.
The DE component consists of magnetic and electric transitions. At
$O(p^4)$ in Chiral Perturbation Theory (ChPT), the magnetic amplitude
consists of a reducible chirally anomalous amplitude~\cite{Ecker}
which is unambiguously determined by the Wess-Zumino-Witten
functional~\cite{WZW} and of a direct anomalous
amplitude~\cite{Bijnens}, arising from chirally covariant odd-parity
octet operators. The direct anomalous amplitude is not calculable in a
model-independent way but the size is expected to be comparable to
that of the reducible amplitude or smaller. At $O(p^6)$ in ChPT,
vector meson exchange (VME) contributions are also possible but, in
contrast to the $K_{L} \rightarrow \pi^+\pi^-\gamma$
decay~\cite{E731DE}, the VME contributions are less important; a weak
deformation model~\cite{WDM} predicts that the VME and the direct
$O(p^6)$ contributions cancel~\cite{Ecker}. There is no definite
prediction from ChPT on the electric transition amplitude, which
depends on undetermined constants. Since the electric amplitude
interferes (INT) with the IB amplitude, it may be distinguished from
the magnetic amplitude, which does not. The experimental determination
of the electric amplitude is of great interest, not only for ChPT,
but for searches for possible non-Standard-Model effects like a
CP-violating asymmetry between $\kpitwog$ and $K^- \! \rightarrow \!
\pi^- \pi^0
\gamma$ decay widths predicted in supersymmetry~\cite{SUSY}.

Previous experiments~\cite{Abrams,Smith,Bolotov} used decay-in-flight
techniques to measure the $\kpitwogam$ branching ratio in the $\pi^+$
kinetic energy region 55 MeV $<$ $T_+$ $<$ 90 MeV with Abrams
$et~al.$~\cite{Abrams} confirming the theoretical prediction for the
IB branching ratio of $2.61\times 10^{-4}$~\cite{DEtheory}.  The
current Particle Data Group average for the DE component of the
branching ratio is $(1.8\pm 0.4)\times 10^{-5}$~\cite{PDG}, which is
five times larger than calculated from the reducible anomalous
amplitude alone and consistent with the coherent combination of that
amplitude and the direct anomalous amplitude suggested by a
factorization model~\cite{Bijnens}.  A decay spectrum shape analysis
of Ref.~\cite{Abrams} indicates the absence of electric contributions
to the DE component but the existing data are not precise enough to
rule out constructive interference with IB.

The differential decay width for the decay $K^+(p) \! \rightarrow \!
\pi^+ (p_{+})\pi^0(p_{0})\gamma(q)$ is conveniently expressed in terms
of a Lorentz invariant variable $W^{2}\equiv (p \cdot q)(p_{+} \cdot
q)/(m_{\pi^+}^{2} m_{K}^{2})$, where $p$, $p_{+}$ and $q$ are
4-momenta of the $K^+$, $\pi^+$ and $\gamma$, and $m_{\pi^+}$ and
$m_{K}$ are masses of $\pi^+$ and $K^+$, respectively. 
The differential width can be written in terms of the IB
differential width as
\begin{eqnarray}
\frac{\partial^2 \Gamma}{\partial T_+ \partial W} =
\frac{\partial^2 \Gamma_{IB}}{\partial T_+ \partial W}
\left[ 1 + 2\frac{m^2_{\pi^+}}{m_K} Re \left( \frac{E}{e A}\right) W^2
+ \frac{m^2_{\pi^+}}{m_K^2} \left( \left|\frac{E}{e A}\right|^2 +
\left|\frac{M}{e A}\right|^2 \right) W^4 \right],
\end{eqnarray}
where $T_+$ is the kinetic energy of the charged pion in the kaon rest
frame, $A$ is the decay amplitude for $K^+\to\pi^+\pi^0$, and $E$ and
$M$ are electric and magnetic DE amplitudes, respectively.  The direct
emission amplitudes $E$ and $M$ are independent of $W$ and their
effects are detectable only in the large $W$ region, while the IB
decay amplitude is proportional to $1/W^2$, and thus dominates in the
small $W$ region.  Experimental determination of these amplitudes
requires high sensitivity to deviations of the measured spectrum from
that of IB over a wide $W$ region.

The E787 detector~\cite{E787det} at the Alternating Gradient
Synchrotron (AGS) of Brookhaven National Laboratory (BNL) was used to
study $K_{\pi2\gamma}$.  Kaons of 790 MeV/$c$~\cite{LESB3} were
incident on a solenoidal spectrometer with a 1.0 Tesla field at a rate
of $7 \times 10^6$ per 1.6-s spill of the AGS. The spectrometer
contains beam detectors, a scintillating-fiber target where the kaons
stop and decay, a central cylindrical drift chamber, and a range stack
(RS) of plastic scintillators with embedded straw chambers.
Measurements of momentum, kinetic energy and range
of charged decay products were performed. The output pulse-shapes of
the RS counters were recorded providing good timing accuracy.  A
hermetic calorimeter system consisting of barrel (BL) and endcap (EC)
detectors measured the positions and energies of photons and other
particles from $K^+$ decays.

The BL calorimeter was used to detect the three $\gamma$'s in the
$\kpitwogam$ decay. It consisted of 48 azimuthal sectors and 4 radial
layers of sandwiches of lead (1 mm) and plastic scintillator (5 mm)
sheets, 14 radiation lengths in depth covering a solid angle of about
3$\pi$ sr.  The visible fraction of the energy deposited in this
system was 30\%.  The Z position (along the beam axis) of the BL hits
was obtained using the charge and timing information from phototubes
on both ends of the 2-m long modules.

The trigger for $\kpitwogam$ required a $K^+$ decay at rest, a charged
track in a kinetic energy region between the endpoint of the $K^+ \!
\rightarrow \!
\pi^+\pi^0\pi^0 $ ($K_{\pi 3}$) decay and the peak of the $K_{\pi
2}$ decay, at least three BL clusters, no extra particles in the
RS, and no energy in the EC.  
The total
number of kaon decays at rest available for the $\kpitwogam$ study was
$1.8 \times 10^{11}$, and a total of $1.1\times 10^{7}$ events
survived the trigger.

The energy and direction of the three $\gamma$'s from the $\kpitwogam$
decay were determined from reconstruction of the BL hits and
the $K^+$ decay vertex position in the target.  The $\pi^+$ momentum
vector and the kinetic energy were measured with the target, the drift
chamber, and the RS.  Thirteen observables were available for the
kinematic fit in this analysis.  The relation between range measured
in the RS and momentum measured in the drift chamber was examined to
reject $\mu^+$ backgrounds. Events were also rejected  based on the
presence of any activity other than three $\gamma$'s and a $\pi^+$.

A kinematic fit was applied to the observables with the following six
constraints: total energy-momentum conservation, consistency of the
charged particle's energy and momentum with a pion hypothesis, and
invariant mass of two $\gamma$'s being equal to $m_{\pi^0}$.  Since
there are three possible pairings of the $\gamma$'s to form the
$\pi^0$, incorrect pairing can move IB events into the region of
large $W$ although IB events tend to populate the region of small $W$.

\begin{figure}
\centerline{
\hbox{\psfig{figure=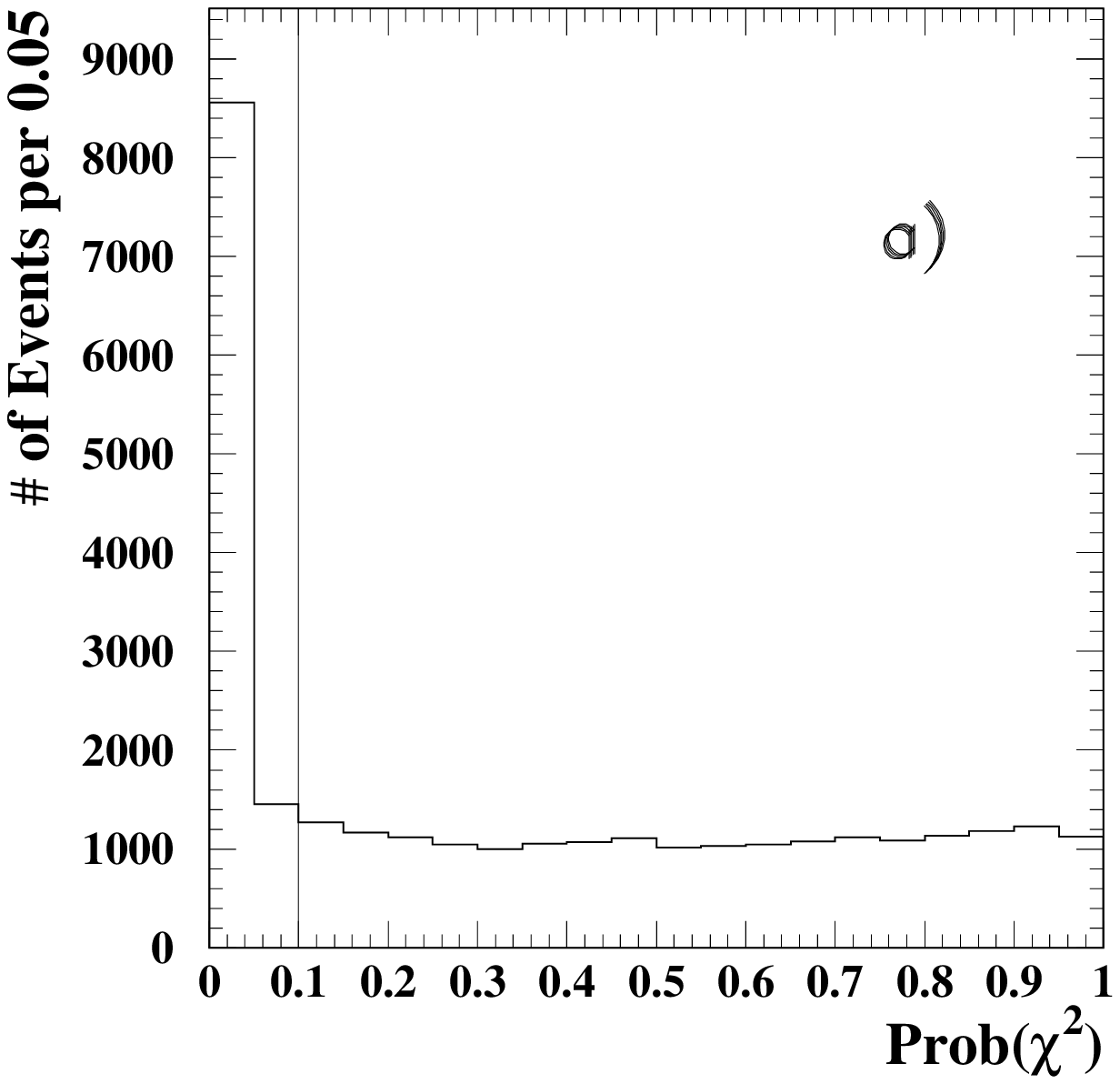,width=8cm}}
\hbox{\psfig{figure=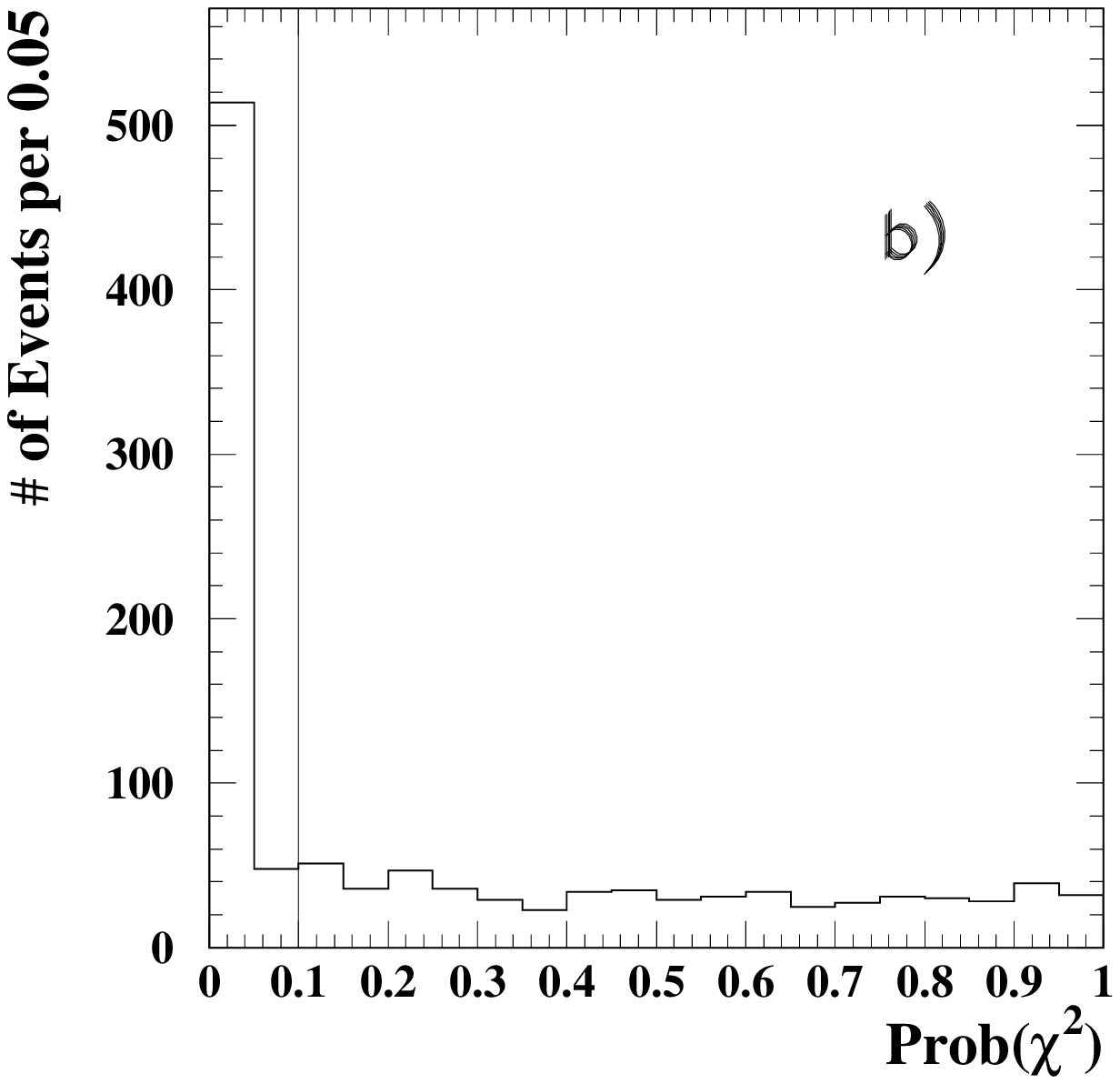,width=8cm}}
}
\caption{
Distribution of the $\chi^2$ probability for the fitted events.
Events with Prob($\chi^2$) $\geq \; 0.1$ are accepted as due to $K^+
\! \rightarrow \! \pi^+ \pi^0 \gamma$ decay. (a) All events. (b)
events in a region of $W >0.5$.  }
\label{figure:probchi}
\end{figure}

To minimize the contamination arising from a wrong combination, the
square of the matrix element of the IB decay was taken as a weight in
the selection of the two $\gamma$'s to be assigned to the $\pi^0$, and
the combination which maximized the product of the $\chi^2$
probability of the kinematic fit, $Prob(\chi^2)$, and the IB weight
was chosen among the three possible event topologies.
Figure~{\ref{figure:probchi}} shows the $\chi^2$ probability
distribution of the fitted events.  The distribution is consistent
with measurement resolution throughout all values of $W$.  
Most of events at $Prob(\chi^2) = 0$ are due to pion nuclear
interactions and muon background, which do not satisfy the kinematic
constraints.

In the Monte Carlo simulation of the IB, DE and INT spectra,
$\kpitwogam$ events were generated with the corresponding matrix
elements in \cite{DEtheory} and analyzed with the code used for real
data.  Thus, the effect of possible incorrect pairing of the two
$\gamma$'s from $\pi^0$ on the Monte Carlo spectra was taken into
account.  The probability of incorrect pairing with this analysis was
estimated to be 2.6\% for IB, 20\% for DE and 9\% for INT. For IB,
2.3\% of the events populate the region of $W>0.5$, and only 6\% of
them were due to incorrect pairing. The estimated $W$ resolution was
about 0.02.

\begin{figure}
\centerline{
\hbox{\psfig{figure=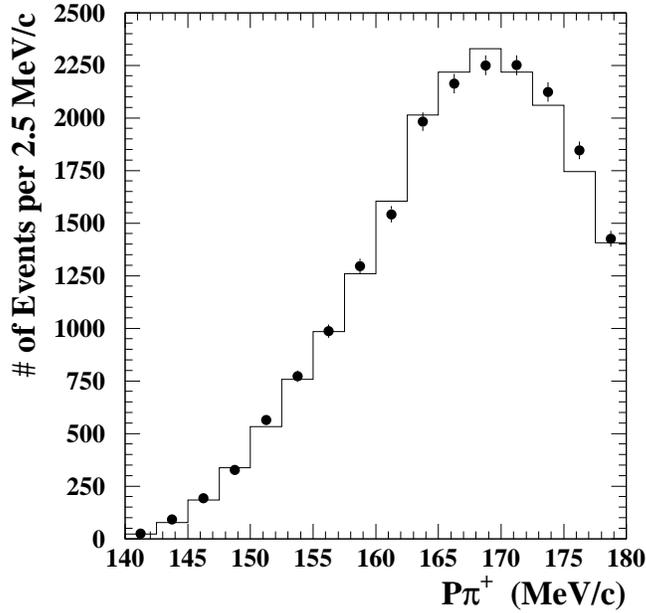,width=10cm}}
}
\caption{
Momentum distribution of observed $\pi^+$ events (dots)
and Monte Carlo simulation with IB alone (solid histogram).
A $\chi^2$ of the fit is 18 for 15 degrees of freedom.  
}
\label{figure:cpimom}
\end{figure}

For the signal selection we required the $\chi^2$ probability
to be more than 10\% and the fitted $\pi^+$ momentum to be between 140
and 180 MeV/$c$; the latter condition on $\pi^+$ momentum was imposed
to remove the $K_{\pi 3}$ and $K_{\pi 2}$ backgrounds as well as
$K^+\to\pi^+\pi^0\gamma$ decays outside the region 55 MeV $<$ $T_+$
$<$ 90 MeV. A total of 19836 events survived all selection cuts,
eight times more than the previous experiments. 
Figure~\ref{figure:cpimom} shows the $\pi^+$ momentum distribution of
the $\kpitwogam$ events after applying all selection cuts.  The
$\pi^+$ momentum distribution is insensitive to the DE contribution,
and thus it is reproduced by a Monte Carlo simulation of the IB
contribution alone.

\begin{figure}
\centerline{
\hbox{\psfig{figure=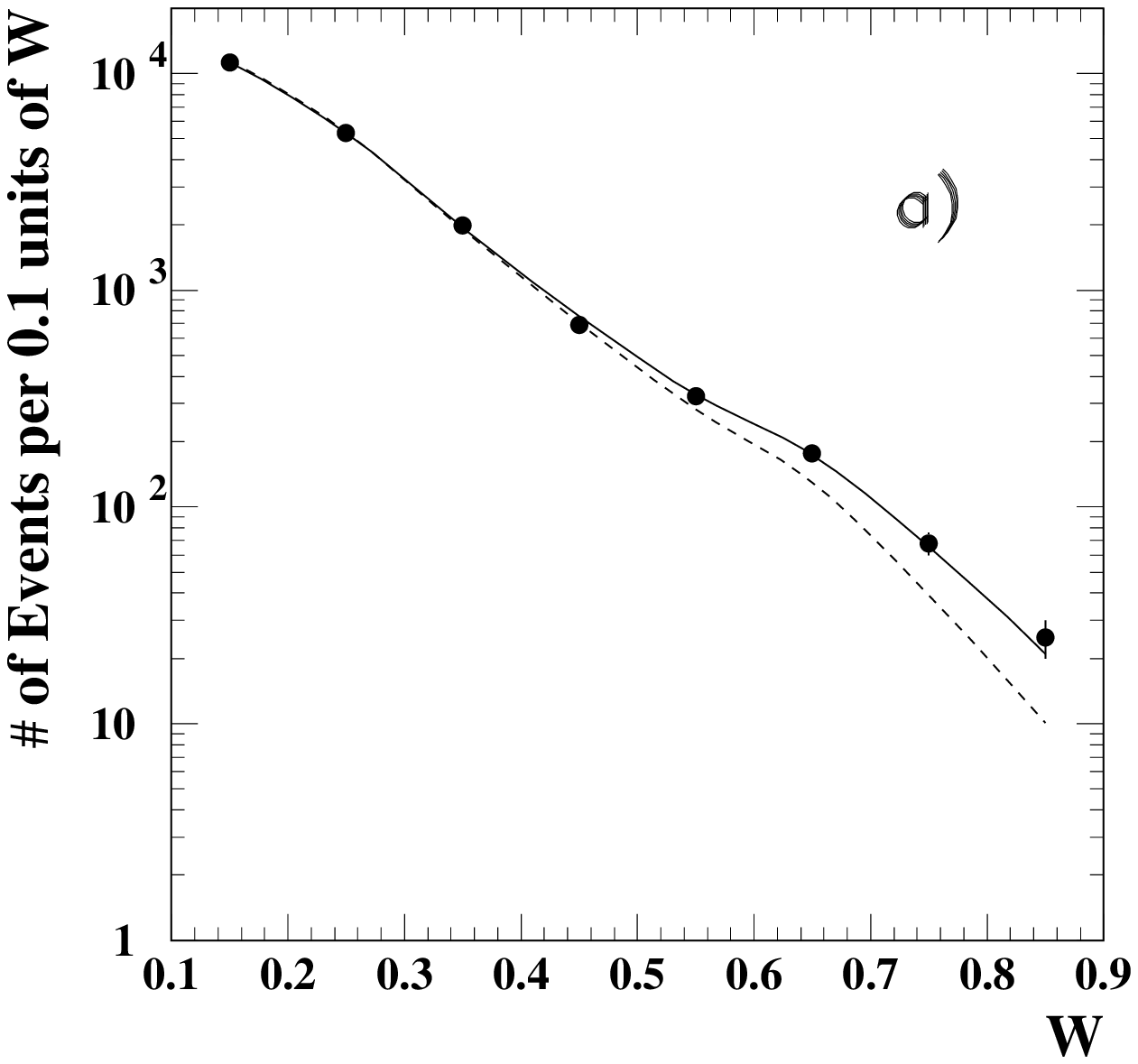,width=8cm}}
\hbox{\psfig{figure=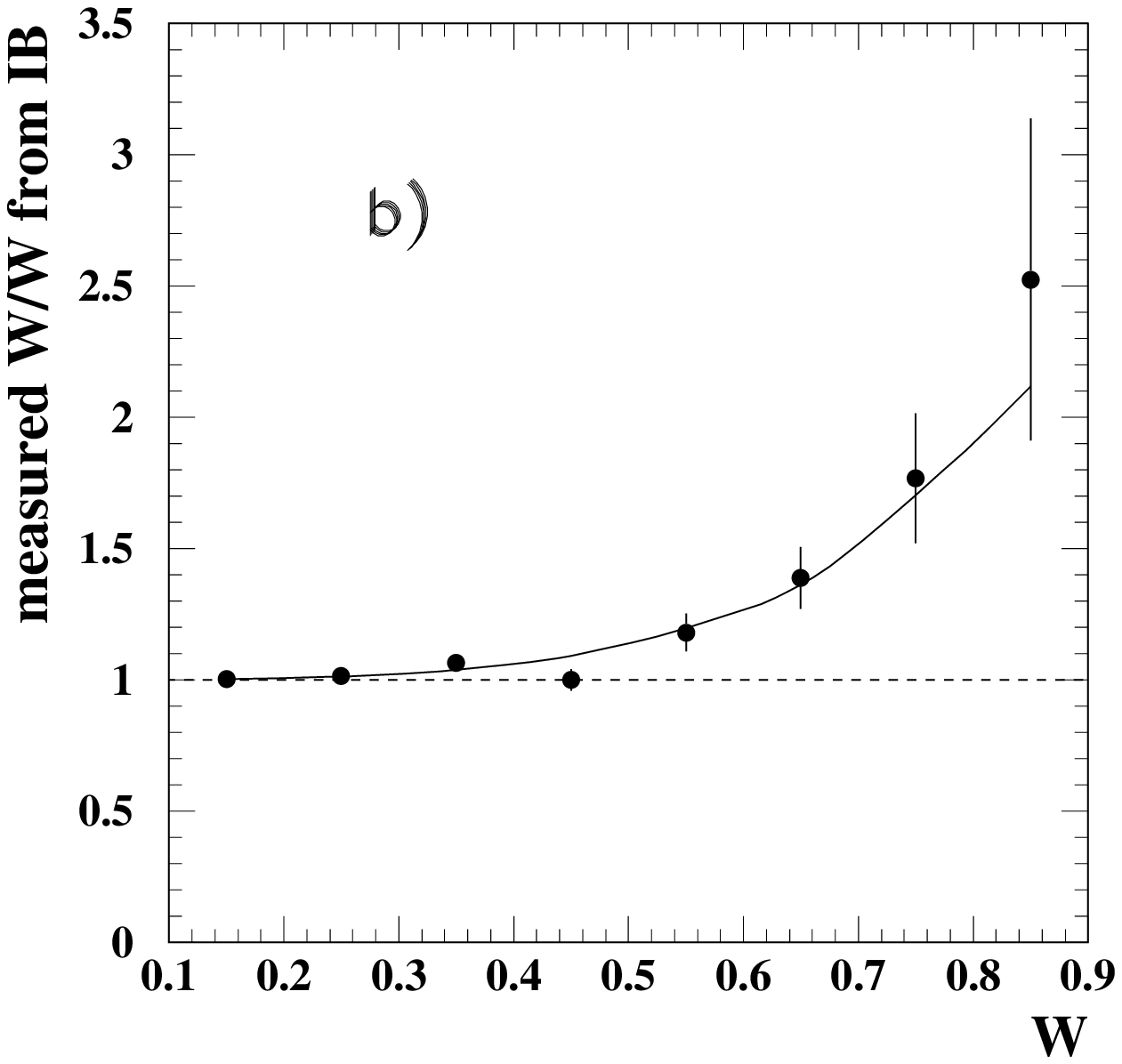,width=8cm}}
}
\caption{a) $W$ spectrum of the observed events 
and best fits to IB+DE (solid curve) and IB alone (dashed curve);  
b) $W$ spectrum normalized to the IB spectrum.}
\label{figure:Wspctl}
\end{figure}

Figure~\ref{figure:Wspctl}a shows the $W$ projection of the signal
events in the $T_+$ signal region.  The dashed curve is a best fit to
IB alone with a $\chisq$ of 61 for 7 degrees of freedom. The solid
curve is the best fit to the sum of IB and DE and gives a $\chi^2$ of
7.9 for 6 degrees of freedom.  Normalized to the IB spectrum the
deviation of the measured spectrum at large $W$ is shown in  
Figure~\ref{figure:Wspctl}b. The DE component was measured to be
$(1.8 \pm 0.3)$\% of the IB component in the region 55 MeV $<$ $T_+$
$<$ 90 MeV.  The branching ratio for DE is determined to be
$B^{T_+}(DE) = (4.7\pm0.8)\times 10^{-6}$ in the $T_+$ signal region
by normalizing to the theoretical prediction for the IB branching
ratio. The best fit to the sum of IB, DE and INT gives the INT
component to be $(-0.4\pm 1.6)$\% of the IB component with a $\chisq$
of 7.8 for 5 degrees of freedom. We conclude that no interference term
is required for the fit to this spectrum.

Several checks to verify the result were made. Spectral analyses were
performed on the $\cos{\theta_{\pi^+ \gamma}}$ distribution, where
$\theta_{\pi^+ \gamma}$ is the opening angle between a $\pi^+$ and a
radiated $\gamma$, and on the energy distribution of the radiated
$\gamma$.  The resulting branching ratios obtained for DE are consistent
with the $W$ fitting result to within 3\%.  This difference is used as
an estimator of the systematic error due to the fitting method.
Selection of $\gamma$ pairs without IB weighting reduced the branching
ratio by 6\% while doubling the statistical uncertainty.  Systematic
uncertainties due to calibration and resolution of the energy and
position measurements were estimated from the observed DE branching
ratio variation as each corresponding Monte Carlo parameter was varied
over an expected range.  The uncertainty due to limited Monte Carlo
statistics was estimated by performing the analysis with small
sub-sets of the Monte Carlo data.  All systematic uncertainties in the
DE branching ratio were added in quadrature to obtain an overall
relative error of $\pm 6$\% ($\pm 0.3 \times 10^{-6}$).
Table~\ref{systematic} gives a summary of the systematic
uncertainties.

Two other possible background sources in the large $W$ region are $K^+
\to \pi^0 \mu^+\nu_{\mu}$ decay with a radiated or accidental $\gamma$, and
$K_{\pi 3}$ decay with a charged pion decay-in-flight.  Both processes
contain a charged muon in the final state.  The estimated number of
the muon background events was 103 in the whole $W$ region (7 out of a
total of 597 with $W > 0.5$), which was obtained by selecting events
with a muon-like range-momentum relation. Imposing additional pion
identification criteria based on the presence of a $\pi^+ \to \mu^+
\nu$ decay signal in the stopping counter~\cite{E787det} did not change 
the $W$ spectrum.  The insensitivity of the final result to the variation of
photon veto cuts also confirmed that these backgrounds are negligible.

Assuming a pure magnetic transition for the DE amplitude, the result
for $B^{T_+}(DE)$ gives $|M| = ( 2.1 \pm 0.2 ) \times 10^{-7}$ for
the dimensionless magnetic amplitude in Eq. (1). The reducible
anomalous amplitude $|- e G_8 m_K^3 / 2
\pi^2 f |$ in Ref.~\cite{Ecker} is $|M| = 1.8 \times 10^{-7}$ with standard
$O(p^2)$ ChPT coupling constants, $f = 93 \; {\rm MeV}$ and $|G_8| =
9\times10^{-6} \; {\rm GeV}^{-2}$. This result supports the
hypothesis that the dominant contribution to DE is due to the reducible
anomalous amplitude and the other magnetic contributions are small or
cancel. Similarly, the null measurement of the INT component gives a
67~\% confidence-level interval of $-2.6 \times 10^{-8} < Re(E) < 1.6
\times 10^{-8}$ for the electric amplitude.

In conclusion, we have performed a new measurement of the
$K^+\to\pi^+\pi^0\gamma$ decay with significantly higher statistics
and improved kinematic constraints, using the E787 detector and $K^+$
decays at rest. The branching ratio for direct emission is $4.7
\pm 0.8 (stat.) \pm 0.3 (sys.) \times 10^{-6}$. The decay spectrum
indicates that the direct emission amplitude is consistent with being
due to purely magnetic contributions comparable to the reducible
anomalous contribution.

We gratefully acknowledge the dedicated effort of the technical staff
supporting this experiment and of the BNL AGS Department.  This
research was supported in part by the U.S. Department of Energy under
Contracts No. DE-AC02-98CH10886, W-7405-ENG-36, and grant
DE-FG02-91ER40671, by the Ministry of Education, Science, Sports and
Culture of Japan through the Japan-U.S. Cooperative Research Program
in High Energy Physics and under the Grant-in-Aids for Scientific
Research, for Encouragement of Young Scientists and for JSPS Fellows,
and by the Natural Sciences and Engineering Research Council and the
National Research Council of Canada.



\begin{table}
\begin{center}
\begin{tabular}{|l|l|}
~Estimated systematic uncertainties& (\%) \\
\hline
~Fitting Method & $\pm 3$ \\
~$\pi^+$ Momentum Calibration &  $^{+3}_{-1}$\\
~$\gamma$ Position Calibration & $^{+3}_{-0}$ \\
~$\gamma$ Position Resolution  & $^{+3}_{-0}$ \\
~$\gamma$ Energy Calibration & $\pm 0.3$ \\
~Monte Carlo Statistics & $\pm 2$ \\ 
~$\gamma$ Interaction in Material & $^{+0}_{-4}$ \\
\hline
Total systematic error & $\pm 6$ \\
\end{tabular}
\end{center}
\caption{\label{systematic}}{
A summary of estimated systematic uncertainties. All uncertainties are
added in quadrature to get the total.}
\end{table}

\end{document}